\documentclass[12pt,fleqn]{article}
\usepackage{verbatim,amssymb,amsmath,graphics}
\renewcommand{\H}{{\cal H}}
\newcommand{\<}{\langle}
\renewcommand{\>}{\rangle}

\newcommand{\bs}{\boldsymbol}

\begin{document}
\renewcommand{\thefootnote}{*}
\title{Rigged Hilbert Space Resonances and Time Asymmetric Quantum 
Mechanics\footnotemark}
\author{A.~Bohm and H.~Kaldass\\University of 
Texas at Austin\\Physics Department\\Austin, TX 78712\\ 
E-mail: bohm@physics.utexas.edu}
\maketitle
\begin{abstract}
The Rigged Hilbert Space (RHS) theory of resonance scattering
and decay is reviewed and contrasted with the standard Hilbert
space (HS) theory of quantum mechanics. The main difference is in the
choice of boundary conditions. Whereas the conventional theory allows
for the in-states $\phi^+$ and the out-states (observables) $\psi^-$
of the $S$-matrix elements $(\psi^-,\phi^+)=(\psi^{out},S \phi^{in})$
any elements of the HS $\H$, $\{\psi^-\}=\{\phi^+\}(=\H)$,
the RHS theory chooses the boundary conditions~:
$\phi^+\in\Phi_-\subset\H\subset\Phi_-^\times$, 
$\psi^-\in\Phi_+\subset \H\subset \Phi_+^\times$, where
$\Phi_-$ ($\Phi_+$) are Hardy class spaces associated to the 
lower (upper) half-plane of the second sheet of the analytically
\footnotetext{Based on a talk delivered at the 
International Conference on Mathematical Physics and Stochastic
Analysis in honor of the 60th birthday of Professor Ludwig Streit.}
continued $S$-matrix. This can be phenomenologically justified
by causality. The two RHS's for states $\phi^+$ and observables $\psi^-$
provide new vectors which are not in $\H$, e.g.\ the 
Dirac-Lippmann-Schwinger kets $|E^{\pm}\>\in\Phi_{\mp}^{\times}$
(solutions of the Lippmann-Schwinger equation with $\pm i\epsilon$
respectively) and the Gamow vectors 
$|E_R-i\Gamma/2^\pm\>\in\Phi_{\mp}^\times$.
The Gamow vectors $|E_R-i\Gamma/2^-\>$ have all the properties that one
heuristically needs for quasistable states. In addition, they give rise
to asymmetric time evolution expressing irreversibility
on the microphysical level.
\end{abstract}
\section{Introduction}
Resonances and decaying states can really not be understood
as autonomous elementary particles in Hilbert space quantum
mechanics because the Hilbert space mathematics does not
allow state vectors characterized by both an energy $E_R$,
and a lifetime $\tau$ (or a width $\Gamma=\hbar/\tau$). This is
in contrast to the way experimentalists analyze their data and list 
their results as Breit-Wigner peak value and width $(E_R,\Gamma)$
for resonances (large values of $\Gamma/E_R$) and as $(E_d,\tau)$
for decaying states if the (Breit-Wigner or Lorentzian) line
shape cannot be resolved but the decay rate can be fitted to an
exponential (small values of $\frac{\hbar}{\tau E_d}$).
Since experimental data are finite in number and there are
always experimental tolerances and interference with 
background, one will never be able to say that the experiment
has precisely established an ideal Breit-Wigner or an exact 
exponential. However, Breit-Wigner energy distribution
and exponential time evolution have been observed so many
times for such an enormous number of different systems in all
areas of physics and chemistry that one can safely take them as
the defining signature of a quasistable state and attribute
observed experimental deviations to 
the ever present background and calculated mathematical
deviations to an unjustified mathematical
idealization. For the mathematical physicists this means that
they have to provide the suitable mathematical idealization.
How and why to do that is the content of our discussion here.
\section{The fundamental calculational tools of quantum mechanics}
In quantum theory one has {\em states} denoted by 
\begin{equation}
\label{1}
\rho\text{ or }W\,,
\text{ or by }\text{$\phi$ for a pure state $\rho=|\phi\>\<\phi|$}\,, 
\end{equation}
and one has {\em observables} denoted by
\begin{equation}
\label{2}
\!\!\!A(=A^{\dagger}),\,\,\,\Lambda,\,\,\, P\,\,(P^{2}=P)\,,
\text{ or by $\psi$, if $P=|\psi\>\<\psi|$, for properties.}
\end{equation}
The vectors $\phi$, $\psi$ form a linear scalar product space $\Phi$.
The operators $\rho$, $W$ and $A$, $\Lambda$, $P$ are linear
operators in $\Phi$. The space $\Phi$ is usually called Hilbert space,
but it is mostly treated like a pre-Hilbert space.
For each ``kind'' of quantum physical system one takes one particular
space $\Phi$.

In an laboratory experiment the state of the quantum 
\renewcommand{\thefootnote}{\arabic{footnote}}
physical system is prepared by a preparation apparatus, e.g.\ an
accelerator. The state $W$ (or $\phi$) is thus experimentally defined 
(``determined'')
by the preparation apparatus. The quantum physical observables are
observed or registered by a registration apparatus, e.g.\ a detector. 
The observable
$\Lambda$ (or $\psi$) or $A$ are thus experimentally defined by the 
registration apparatus.

In experiments with quantum systems one measures ratios of large integers
$N_{i}/N$ or $N(t)/N$, e.g.\ as ratios of detector counts of the $i$-th
detector $N_{i}$ and counts of all detectors $N$ or as ratios of detector
counts $N(t)$ in the time interval between $t=0$ and $t=t$ and in 
``all'' time $N=N(\infty)$. This ratio of large numbers
is interpreted as probability, e.g.\ as probability ${\cal P}(P_i)$
for a property $P_i$
\begin{subequations}
\begin{equation}
\label{3a}
\frac{N_i}{N}\approx {\cal P}(P_{i})
\end{equation}
or as probability for the observable $\Lambda(t)$ at a time $t$
\begin{equation}
\label{3b}
\frac{N(t)}{N}\approx {\cal P}(\Lambda(t))
\end{equation}
\end{subequations}
where the observables $P_i$ or $\Lambda$ are experimentally given
(``defined'') by the detector.
For a more general observable 
\begin{equation}
\label{4}
A=\sum_{i}^{\infty}a_iP_i
\end{equation}
one obtains the average value (of the eigenvalues $a_i$)
\begin{equation}
\tag{3c}
\begin{split}
\sum_{i=1}^{\text{finite}}a_i\frac{N_i}{N}\approx 
\sum_{i=1}^{\infty}a_i{\cal P}(P_i)\\
\sum_{i=1}^{\text{finite}}\frac{N_i}{N}\approx\sum_{i=1}^{\infty}
{\cal P}(P_i)=1\,.
\end{split}
\end{equation}
The symbol $\approx$ denotes the association of the experimentally measured quantity on
the left hand side with the theoretically calculated quantity
on the right hand side.

In quantum theory the probability of an observable $\Lambda$ in the 
state $W$ at time $t$ is calculated as
\begin{subequations}
\label{5}
\begin{equation} \label{5a}
{\cal P}(t)={\cal P}(\Lambda(t))={\rm Tr}(\Lambda(t)W_{0})
={\rm Tr}(\Lambda_{0}W(t))\,.
\end{equation}
If the state is pure, given by the state vector $\phi$, and if the 
observable is a property given by the one-dimensional projector
$|\psi\>\<\psi|$, or given by the ``observable'' vector $\psi$, then
\begin{equation} \label{5b}
{\cal P}(t)=|\<\psi|\phi(t)\>|^{2}=|\<\psi(t)|\phi\>|^{2}\,.
\end{equation}
\end{subequations}
The trace in~\eqref{5} is calculated using any basis system of the 
space $\Phi$; either any discrete basis 
\begin{subequations}
\label{6}
\begin{equation} \label{6a}
\Phi\ni\phi=\sum_{i}|i\>\<i|\phi\>\, ;
\end{equation}
or any continuous basis (Dirac basis vector expansion)
\begin{equation} \label{6b}
\Phi\ni\phi=\int d\lambda|\lambda\>\<\lambda|\phi\>
\end{equation}
\end{subequations}
or any basis system consisting of discrete and continuous
(generalized) eigenvectors of a complete system of commuting 
observables. Thus the trace is given by e.g.\
\begin{subequations}
\label{7}
\begin{equation} \label{7a}
{\rm Tr}(\Lambda W)=\sum_{i}\<i|\Lambda W|i\>\,\,\text{ or }\,\,
{\rm Tr}(\Lambda W)=\int d\lambda\<\lambda|\Lambda W|\lambda\>\,.
\end{equation}
For the special case~\eqref{5b} the probability for the 
observably $\psi$ in the state $\phi$ is given by
\begin{align} 
\label{7b}
{\rm Tr}(|\psi\>\<\psi|\phi\>\<\phi|)&=|\<\psi|\phi\>|^{2}
=\left|\sum_{i=0}^{\infty}\<\psi|i\>\<i|\phi\>\right|^{2}\,,\\
\nonumber \text{or by}&\\
\label{7c}
&=\left|\int d\lambda\<\psi|\lambda\>\<\lambda|\phi\>\right|^{2}\,
\end{align}
\end{subequations}
if one uses a continuous basis of Dirac kets $|\lambda\>$.

The time evolution in~\eqref{5a} and~\eqref{5b} (dynamics of the 
quantum system) is given by the Hamiltonian operator $H$ of the system;
either as
\begin{align} 
\tag{8b}
&\frac{\partial W}{\partial t}=\frac{i}{\hbar}[H, W(t)]\qquad(8a)\,;&
\quad i\hbar\frac{\partial \phi(t)}{\partial t}=H\phi(t)\\
\nonumber
&\quad\phi(t=0)=\phi_{0}\,(\text{ Schroedinger picture})\, ,& \\
&\nonumber\text{or by~:}& \\
\tag{8d}
&\frac{\partial\Lambda}{\partial t}
=-\frac{i}{\hbar}[H,\Lambda(t)]\qquad(8c)\,;&
\quad i\hbar\frac{\partial \psi}{\partial t}=-H\psi(t)\,.\\
\nonumber
& \,\,\,\,\,(\text{Heisenberg picture})
\end{align}
\addtocounter{equation}{1}
None of the above equations is mathematically precise until we define the space $\Phi$,
the kets $|\lambda\>$ or the integration $d\lambda$ 
in~\eqref{6b} and~\eqref{7c}, and
specify the initial-boundary conditions $\phi_0$ etc.\ for the equations~$(8)$.
Before we choose these mathematical definitions we want to
discuss some phenomenological properties of the above quantities which 
may influence these choices.
\section{Empirical reasons for time asymmetric quantum mechanics}
The time $t$ in~\eqref{5} and in the operators $\Lambda(t)$
and $W(t)$ is usually allowed to take positive and negative values, i.e.\ 
one chooses time symmetric boundary conditions for the Schroedinger
and Heisenberg equations of motion~$(8)$. This choice is not
compelling from an empirical point of view
for the following reason.

An obvious expression of causality is that a state $W_0$ (or $\phi_0$)
has to be prepared first {\em before} an observable $\Lambda(t)$
or $\psi(t)$ can be measured in it.
If one calls $t=0$ the time at which the preparation of the state $W$ 
(or $\phi$)
is completed then the time translation in $\Lambda(t)$ (or $\psi(t)$)
in~\eqref{5} makes physically sense only for $t\geq 0$. The registration
counts $N(t)$ in~\eqref{3b}, by which ${\cal P}(t)$ is measured, can be
taken in the future $t>0$, but not in the past. Thus, ${\rm Tr}(\Lambda(t)W)$
will have an experimental counterpart for $t\geq 0$ but not for $t<0$. The
experimental data for ${\cal P}(t)$ would give that
\begin{subequations}
\label{10}
\begin{equation} \label{10a}
{\cal P}(t)={\rm Tr}(\Lambda(t)W)\approx\frac{N(t)}{N}\,\,\text{ for }t>0
\end{equation}
and
\begin{equation} \label{10b}
{\cal P}(t)={\rm Tr}(\Lambda(t)W)\approx 0\,\,\text{ for }t<0
\end{equation}
\end{subequations}
because if the detector would click before the state is prepared we 
would discount this click as noise.

Often, such as for stationary states and/or time independent observables,
it does not matter at which time ${\cal P}(t)={\rm Tr}(\Lambda(t)W)
={\rm Tr}(\Lambda W(t))$ is calculated. But one cannot make it a general
principle that time evolution of the observable $\Lambda(t)$ (or of the 
state $W(t)$) must go into both directions. In fact the most natural
description of the experimental situation~\eqref{10} would be to
require a time ordering in the probability formula
\begin{equation}
\label{11}
{\cal P}(t)={\rm Tr}(\Lambda(t)W(0))\quad t>t_0=0
\end{equation}
and admit in the mathematical description only time translations of
observables $\Lambda(t)$ relative to the state $W(t_0)$ by an amount
$t-t_0>0$. This would not only reflect the experimental
situation, which forbids time translation of the registration apparatus 
relative
to the preparation apparatus to a time $t<t_0$. It would
also incorporate the notion of {\em causality} into the mathematical theory
on the operator level.

The time translation is quite different from other transformations of the 
space-time symmetry group, e.g.\ rotations and space-translations. 
These symmetry 
transformations of space-time are experimentally realized as 
transformations
of the registration apparatus (detector) relative to the preparation apparatus
(accelerator). Whereas space translations and rotations of the apparatuses 
can be
performed back and forth, the time at which the detector is activated
can only be shifted into the future not backward past the time
of preparation. Thus time
translations of the apparatuses form only a {\em semigroup} whereas rotations
and translations of the apparatuses form a group. Therefore it is natural
to represent space translations and rotations by unitary groups 
of operators in the space of physical states. But it is unjustified
to assume that time translations are also always represented by group operators
in the space of states, i.e.\ to assume that in quantum mechanics only states
with reversible time evolution exist.

When we make the choices that give  a precise mathematical meaning to the 
vectors $\phi$ and $\psi$ and the operators $W$, $\Lambda$, $A$, $A^{\dagger}$
in~\eqref{1} and~\eqref{2} we should therefore not restrict ourselves
to those mathematical assumptions that dictate  a unitary group evolution.
\section{The Hilbert space idealization has reversible time evolution}
Of the vectors $\phi$, $\psi$ we have so far assumed that they fulfill the
mathematical axioms of a linear space with scalar product $\<\phi|\psi\>$
and that $W$, $\Lambda$, ... are linear operators in this space. 
In addition, we needed some rules like~\eqref{6} that permit the 
calculations in~\eqref{7}; we have not yet said anything about the 
convergence of infinite sequences and the meaning of integration in~\eqref{7}.
In order to prove existence theorems mathematicians need their spaces to be
complete (topologically)---they use spaces that also fulfill
topological axioms. This gives infinite series, like the ones that
occur in~\eqref{6},~\eqref{7} and~\eqref{4} and on the right hand side 
of~$(3\rm{c})$,
a well defined meaning of convergence. It also defines the meaning of 
such continuous operator functions of a parameter $t$ as occur 
on the right hand side of~\eqref{5} and~\eqref{3b}. These topological properties
(the definition of open sets) cannot be directly inferred from experimental
data as one can see by comparing the right hand side (calculated theoretical quantities) of~\eqref{3b} and~$(3\rm{c})$  with the left hand side of~\eqref{3b}
and~$(3\rm{c})$ (measured experimental quantities).
The left hand side of~\eqref{3b} changes as a function of $t$ only in integer
steps of $\frac{1}{N}$, whereas the right hand side is presumed to be 
a continuous function of time. Similarly, the experimental expression
on the left hand side of~$(3\rm{c})$ has a finite sum, in the theoretical
expression on the right hand side the sum is infinite. Experiments necessarily involve
a finite amount of data. Measured probabilities therefore are only
approximate probabilities and cannot supply continuous functions.
They cannot determine the meaning of continuous operator functions
(like $\Lambda(t)$) or continuous vector functions (like $\psi(t)$) or 
the meaning of convergence of infinite sequences 
as in~$(3\rm{c})$ or in~\eqref{7a} or other topological notions.
The equality $\approx$ between experimental and mathematical
quantities has a meaning only within certain experimental errors
and for large numbers $N$.
Therefore, one has 
to make some---more or less---arbitrary mathematical idealizations
in order to obtain complete mathematical structures, like linear topological
spaces and topological algebras of operators.

One such idealization is the Hilbert space (HS). 
This is obtained by adjoining
to the linear scalar product space $\Phi$ the limit element of infinite
converging (Cauchy) sequences, with the Hilbert space convergence
(or Hilbert space topology $\tau_{\H}$) defined by
\begin{equation}
\label{0}
\phi_{\nu}\stackrel{\tau_{\H}}{\rightarrow}\phi\quad\text{iff}\quad\|\phi_{\nu}-\phi\|
\rightarrow 0\,\,
\text{ for }\nu\rightarrow \infty
\end{equation}
where the norm $\|\phi\|$ is defined by the scalar product
$\|\phi\|\equiv\sqrt{(\phi,\phi)}$. Thus the infinite
sequences of the discrete components $\{\<i|\psi\>\}_{i=1}^{\infty}$,
$\{\<i|\phi\>\}_{i=1}^{\infty}$ of the vectors $\psi\,,\phi\in\H$
in~\eqref{6b} are square summable. The continuous components (wave-functions)
$\{\<\lambda|\psi\>=\psi(\lambda)\,\,|\,\,\lambda\in
\text{Spectrum of self-adjoint operator}\}$ and 
$\{\<\lambda|\phi\>=\phi(\lambda)\,\,|\,\,\lambda\in\text{ Spectrum}\}$
are square integrable functions. However, the integrals 
\begin{equation}
\label{14}
(\psi,\phi)=\int d\lambda\overline{\<\lambda|\psi\>}\<\lambda|\phi\>
\end{equation}
are not Riemann but are Lebesgue integrals. This means the values of the 
functions $\psi(\lambda)=\<\lambda|\psi\>$ at a particular point
(or at all rational numbers) are not defined, which in turn means
that the Dirac kets $|\lambda\>$ cannot be defined. The Lebesgue integration
also makes the interpretation of the probability density
$|\<E|\psi\>|^{2}$ as the energy resolution of the detector
for the observable $|\psi\>\<\psi|$ rather unintuitive, at least for
those $\psi\in\H$ that do not have a smooth function in the class of Lebesgue
integrable functions $\{\<E|\psi\>\}$ which represent $\psi$~\cite{7}.

In the (complete) Hilbert space one can define self-adjointness and give the 
precise meaning of ``spectral resolution'' to equations like~\eqref{4}. One can
make the hypothesis that observables are (not just hermitian or symmetric
but) self-adjoint operators and one can prove existence theorems.

One of these existence theorems is the Gleason theorem~\cite{8} 
which states that
if ${\cal P}(P_i)$ is the function on the set of projectors $\{P_i\}$ which
fulfills the axioms of probabilities~\footnote{For ${\cal P}(P_i)$ or
${\cal P}(\Lambda)$ to be a probability it has to fulfill the axioms of 
probability theory~:
$${\cal P}(P)\geq 0\,,\quad{\cal P}(1)=1\,,\quad{\cal P}(P_i+P_j)=
{\cal P}(P_i)+{\cal P}(P_j)\,,$$
$$\left([P_i,P_j]=0\,,\,P_i P_j=0\right)\,.$$}
\addtocounter{footnote}{-1}then there exists a positive
trace class operator (density operator) $\rho$
in $\H$ such that ${\cal P}(P_i)={\rm Tr}(P_i\rho)={\cal P}_{\rho}(P_i)$.
This operator $\rho$ thus defines the state. Since the converse 
(for any positive trace class operator 
$\rho$, ${\cal P}_{\rho}(P_i)\equiv{\rm Tr}(P_i\rho)$
fulfills the axioms of probability theory~\footnotemark) is simple to see, 
one was led to the conclusion~\cite{9} 
that there is one to one correspondence
between 
\begin{subequations}
\label{15}
\begin{equation} \label{15a}
\text{quantum physical states }\Leftrightarrow\text{ density operators $\rho$}
\end{equation}
and in particular between
\begin{equation} \label{15b}
\!\!\!\!\!\text{pure quantum physical states }
\Leftrightarrow\text{ vectors $\phi$ (up to a phase) in
$\H$}
\end{equation}
\end{subequations}

Another existence theorem (Stone-von Neumann operator calculus~\cite{10}) 
asserts that the Cauchy problem in quantum mechanics, e.g.\ in the 
form of the Schroedinger equation~$(8\rm{b})$ with the initial condition
$\phi(t=0)=\phi_0\in\H$ and a selfadjoint Hamiltonian $H$ with domain
$D(H)$ dense in $\H$, has a solution given by the one parameter
group of (strongly continuous) unitary operators $U^{\dagger}(t)$
in $\H$~:~\footnote{The operator $U^{\dagger}(t)$ is defined in all
of $\H$ by the Stone-von Neumann calculus 
as $U^{\dagger}(t)=\int_{E_{0}}^{\infty}e^{-iEt}dP(E)$
not by the exponential series
$$
e^{-iHt}=\sum_{n=0}^{\infty}\frac{t^{n}}{n!}(-iH)^{n}
$$
which is defined (converges) only on a dense subspace 
${\cal A}\subset D\subset \H$ called analytic vectors.}
\begin{equation}
\tag{14a} 
\phi(t)=U^{\dagger}(t)\phi_0\equiv e^{-iHt}\phi_{0}\quad
-\infty<t<\infty
\end{equation}
\addtocounter{equation}{1}
where $\phi(t)$ depends continuously on the initial condition.
This means
\begin{equation} \label{17}
U^{\dagger}(t+\tau)=U^{\dagger}(t)U^{\dagger}(\tau)\,,
\quad U^{\dagger}(-t)=(U^{\dagger})^{-1}(t)=U(t)
\end{equation}
\begin{equation} \label{18}
\frac{d^{p}U(t)}{dt^{p}}\phi=(-iH)^{p}U^{\dagger}(t)\phi
=U^{\dagger}(t)(-iH)^{p}\phi\quad\text{for }\phi\in D(H)\,.
\end{equation}
The same result holds for the time evolution of the observable
vector $\psi$ in $\H$
\begin{equation} 
\tag{14b}
\psi(t)=U(t)\psi_0=e^{iHt}\psi_0\quad -\infty<t<\infty\,.
\end{equation}
%\addtocounter{equation}{2}
In terms of the operators $W$ and $\Lambda$ this is given by
\begin{subequations}
\label{21}
\begin{equation} \label{21a}
W(t)=U^{\dagger}(t)W_0 U(t)=e^{-iHt}W_{0}e^{iHt}\quad -\infty<t<+\infty
\end{equation}
\begin{equation} \label{21b}
\Lambda(t)=e^{iHt}\Lambda_{0}e^{-iHt}\quad -\infty<t<+\infty\,.
\end{equation}
\end{subequations}

The conclusion that one draws from these two existence theorems
is the following~:\\
The choice to describe states in the HS theory is very restricted
and can only be given by a density operator (positive definite
trace-class) and (for a pure state) by a vector $\phi\in\H$. 
The time evolution of these states must be the time symmetric
reversible group evolution~$(14)$. There are {\em no} states in the 
HS quantum mechanics of closed systems (which 
fulfill~$(8\rm{a})$ or~$(8\rm{b})$) 
which have {\it asymmetric} (irreversible) 
time evolution~\footnote{Irreversibility in conventional
quantum theory is considered to be ``non-quantum mechanical'' and
always thought of as being due to external influences upon ``open''
systems. It is described by an additional term on the right hand side of
$(8\rm{a})$ which is not given by the commutator with $H$ of the quantum
system.}.
The evolution of the quantum mechanical observables is also
time symmetric, i.e.\ the observables $\Lambda'(t-t_0)$ can evolve
relative to the state $W(t_0)$ by an arbitrary amount $t-t_0\geq 0$
as well as $t-t_0\leq 0$. For every $U(t_2-t_1)$ that time translates
the observable $\psi(t_{1})$ to the observable $\psi(t_{2})$
there exists also a time translation $U(t_1-t_2)=U^{-1}(t_2-t_1)$
that reverses the transformation.

The time symmetric dynamics~\eqref{21a} and~\eqref{21b} that is a
mathematical consequence of the Hilbert space idealization
(topology $\tau_{\H}$ defined by~\eqref{0}) contradicts the phenomenological 
time ordering for the observed probabilities~\eqref{11}. 
This time ordering is expressed by the 
preparation $\rightarrow$ registration arrow of time~: a state 
must be prepared first before an observable can be measured
in it. It expresses causality in
quantum mechanics. Quantum theory in HS
is time symmetric and cannot express causality. 
A consequence of this result is in particular that in the HS theory there
cannot exist a state vector $\phi^{D}$ (or more general state $W^{D}$)
which has been created at any finite time $t_0\ne -\infty$
(which we call $t_0=0$) and then decays into decay products.

This mathematical consequence is not
surprising if one keeps in mind that in the HS theory
the $t$-evolution--where $t$ is the relative time between state
and observable--is given by a reversible unitary group. One
cannot just impose on the group evolution an arbitrary condition like
causality, chopping off one half of the theory,
and expect that what remains is still a consistent theory.

The reversibility of quantum mechanics in HS and the violation
of the causality principle is a  consequence of the mathematical
idealization given by the topology of the HS, i.e.\ by~\eqref{0}.
It is not a consequence of the fundamental hypothesis of
quantum physics as given by the interpretation~\eqref{5}
and by the dynamical equation~$(8)$. We shall therefore
return to the quantum mechanical Cauchy problem~$(8)$
and modify the boundary conditions $\phi_{0}\in\H$ such that
also semigroup time evolution is possible. Mathematically,
the simplest modification
is to go to strongly continuous semigroups in a Banach space.
But since we have already a scalar product $(\,\,\, ,\,\,\, )$ in the 
linear space $\Phi$ (and we need the scalar product to calculate
such physical quantities like the probabilities $|(\phi,\psi)|^{2}$), 
Banach space completion is
no more an option.
\section{The Rigged Hilbert Space idealization has irreversible
semigroup evolution and Gamow vectors with exponential decay}
We shall complete the linear scalar product space $\Phi$ into a 
locally convex nuclear space with a topology stronger than the
Hilbert space topology given by the scalar product. Specifically
we shall choose a countable Hilbert space where the meaning 
of convergence, i.e.\ the topology $\tau_{\Phi}$ is defined by
a countable number of scalar products $(\,\,\, ,\,\,\, )_{p}$, 
$p=0,1,2,\ldots$ where $(\,\,\, ,\,\,\,)_{p=0}=(\,\,\, ,\,\,\,)$
is the scalar product of the HS. 
Convergence with respect to $\tau_{\Phi}$,
$\phi_{\nu}\stackrel{\tau_{\Phi}}{\rightarrow}\phi$, means~:
\begin{equation} \label{19}
\begin{split}
\phi_{\nu}\stackrel{\tau_{\Phi}}{\rightarrow}\phi
\text{ iff }\|\phi_{\nu}-\phi\|_{p}^{2}&=(\phi_{\nu}-\phi,\phi_{\nu}-\phi)_{p}
\rightarrow 0\text{ for }\nu\rightarrow \infty\\
&\text{ for every }p=0,1,2,\ldots\,.
\end{split}
\end{equation}
This topology is stronger than $\tau_{\H}$. 
The countable number of scalar products, i.e.\ the topology
$\tau_\Phi$, is usually chosen such that the algebra of
observables for the physical system under consideration
becomes an algebra of $\tau_\Phi$-continuous operators.
If we denote
the $\tau_{\Phi}$-completion of the linear space $\Phi$
again by $\Phi$ then we have the two complete topological spaces
$\Phi\subset\H$ with $\Phi$ dense in $\H$.
Taking in addition the space $\Phi^{\times}$ of 
$\tau_{\Phi}$-continuous antilinear functionals $F(\phi)$ on $\Phi$, 
and the space
$\H^{\times}$ of $\tau_{\H}$-continuous functionals
$f(h)=(h,f)$ which are given by the scalar product, we 
obtain the Gelfand triplet or Rigged Hilbert Space (RHS)~\cite{11}
\begin{equation}
\label{23}
\Phi\subset\H=\H^{\times}\subset\Phi^{\times}\,.
\end{equation}
We shall use the Dirac notation for the $\tau_{\Phi}$-continuous
functionals $F(\phi)\equiv\<\phi|F\>$, because
$F(\phi)$ is an extension of the scalar product
$(h,f)$ to those $F\in\Phi^{\times}$ which are not in $\H$.

In these RHS's (one for each kind of quantum physical system)
Dirac's formalism of kets (with a continuous set of eigenvalues)
and the continuous basis vector expansion~\eqref{6b}
attain a mathematical meaning and the integrals in~\eqref{7}
are Riemann integrals.

These RHS's also allow for time-asymmetric
solutions of the quantum mechanical Cauchy problem~$(8)$.

In the RHS formulation one can choose different subspaces of
$\H$ to distinguish between states and observables.
We call $\Phi_-$ the space that describes the states (called in-states
in the scattering experiment) prepared by
preparation apparatuses (e.g.\ accelerator). We call $\Phi_+$
the space that describes the observables (called out-states
in scattering theory)
registered by the registration apparatuses (e.g.\ detector).
The two subspaces $\Phi_-$ and $\Phi_+$ are not disjoint.
The HS formulation, in
contrast, does not allow for this mathematical distinction
into separate subspaces of states and observables.
Thus there is one Hilbert space $\H$ and for each quantum mechanical 
(scattering) system two dense subspaces $\Phi_\mp$ and therewith
two RHS's
\begin{subequations}
\label{24}
\begin{align}
\label{24a}
&\!\!\!\!\!\!\!\!
\Phi_-\subset\H\subset\Phi_-^\times\text{ with the physical 
interpretation as in-states and}\\
\label{24b}
&\!\!\!\!\!\!\!\!
\Phi_+\subset\H\subset\Phi_+^\times\text{ with the physical
interpretation as out-observables}\,.
\end{align}
\end{subequations}
Mathematically $\Phi_{\mp}$ are defined by their realization
as function spaces for their energy wavefunctions~:
\begin{subequations}
\label{25}
\begin{align}
\label{25a}
\phi^{+}\in\Phi_-\Leftrightarrow\<^{+}E|\phi^+\>\in\left.{\cal S}
\cap\H_-^2\right|_{{\mathbb R}^{+}}\\
\label{25b}
\psi^-\in\Phi_+\Leftrightarrow\<^-E|\psi^-\>\in\left. {\cal S}
\cap \H_{+}^{2}\right|_{{\mathbb R}^{+}}
\end{align}
\end{subequations}
where ${\cal S}$ denotes the Schwartz space of functions
and ${\cal H}_-^2$, ${\cal H}_+^2$ denotes Hardy class functions
in the lower and upper complex plane, respectively. 
(Here complex half-planes refer to the second Riemann sheet
of the $S$-matrix). The $^{\pm}$ in $\<^{\pm}E|$ refers to the 
$\pm i\epsilon$ in the Lippmann-Schwinger equation for the 
eigenkets of $H=H_0+V$.

The interpretation~\eqref{24} of the mathematical
spaces~\eqref{25} can be inferred from the 
preparation $\rightarrow$ registration arrow of time~\cite{12}.

In addition to the vectors $\phi^+\in\Phi_-\subset\H$ and the 
$\psi^-\in\Phi_+\subset\H$ defined by the experimental apparatus,
the RHS formulation also provides elements outside of $\H$,
e.g.\ Dirac's scattering states 
$|\bs{p}\>=|E,\theta_p,\phi_p^{\mp}\>\in\Phi^{\times}_{\pm}$ and the 
Gamow (decaying) states (or Gamow kets) 
$\psi^{G}=|E_R-i\Gamma/2^{-}\>\in\Phi_+^\times$. A Gamow ket
is a generalized eigenvector of a self-adjoint
Hamiltonian extended to $\Phi^{\times}$, with complex
eigenvalue $z_R=E_R-i\Gamma/2$, precisely,
\begin{equation}
\label{26}
\begin{split}
\<H\psi^-|E_R-i\Gamma/2^-\>&\equiv\<\psi^-|H^\times|E_R-i\Gamma/2^-\>\\
&=(E_R-i\Gamma/2)\<\psi^-|E_R-i\Gamma/2^-\>
\end{split}
\end{equation}
for all $\psi^-\in\Phi_{+}$ (the space of observed decay products).
Here $E_R$ represents the resonance energy and $\Gamma$ 
the width of the Breit-Wigner energy distribution.
The superscript $^-$ in $|E_R-i\Gamma/2^-\>$ is inherited
from the kets of the Lippmann-Schwinger equation,
the subscripts on the corresponding space $\Phi_+^\times$ 
from the mathematicians' notation for Hardy class functions~\eqref{25}.
Scattering theory in physics and Hardy class functions in
mathematics were developed independently of 
each other. 
Except for this discrepancy in the notation for the 
labels $\mp$, the Hardy
class spaces $\Phi_-$ provide an excellent
mathematical 
image for prepared states $\{\phi^{+}\}$ and the spaces
$\Phi_+$ for the registered
observables $\{\psi^-\}$ in the quantum theory of scattering and
decay. This match is a wonderful example of what Wigner
called ``the miracle of the appropriateness of the language
of mathematics for the formulation~... of physics''~\cite{13}.

The Gamow kets have asymmetric time evolution that obeys an
exact exponential law~\cite{4}
\begin{subequations}
\label{27}
\begin{equation}
\label{27a}
\begin{split}
\psi^{G}(t)=e^{-iH^{\times}t}_{+}|E_R-&i\Gamma/2^-\>
=e^{-iE_R t}e^{-\Gamma t/2}
|E_R-i\Gamma/2^-\>\,,\\
&\text{for }t\geq 0\text{ only}\,.
\end{split}
\end{equation}
The Gamow kets are solutions of the Schroedinger equation~$(8\rm{b})$
but do not fulfill the Hilbert space boundary condition.
Instead they fulfill the time asymmetric boundary condition
$\psi^{G}(t=0)\in\Phi_{+}^{\times}$. Whereas the first part
of~\eqref{27a} can be formally verified from~\eqref{26},
the derivation of the time asymmetry $t\geq 0$ is 
highly non-trivial and requires specific properties of
Hardy class functions.
The semigroup $e^{-iH^{\times}t}_{+}$ is only defined for positive 
values of the time, $t\geq 0$, ($H^\times$ is the operator
$H^\dagger$ extended into $\Phi^{\times}$ defined by the first equality
of~\eqref{26}).
There is another semigroup $e^{-iH^{\times}t}_{-}$, $t\leq 0$
and another Gamow ket $\tilde{\psi}^{G}=|E_R+i\Gamma/2^+\>\in\Phi_-^\times$
with the asymmetric time evolution
\begin{equation}
\label{27b}
\begin{split}
\tilde{\psi}^{G}(t)=e^{-iH^\times t}_{-}|E_R+&i\Gamma/2^+\>
=e^{-iE_R t}e^{+\Gamma/2 t}|E_R+i\Gamma/2^+\>\,,\\
&\text{for }t\leq 0\text{ only}\,.
\end{split}
\end{equation}
\end{subequations}
The Gamow vectors are defined from the pole term of  
the analytically continued $S$-matrix
at the resonance position 
at $z_R=E_R-i\Gamma/2$ (and at $z_R=E_R+i\Gamma/2$ for~\eqref{27b})
in the second Riemann sheet.
From this one obtains, using the Hardy class property,
the Breit-Wigner energy distribution for their wave function~\cite{4,14}
\begin{equation}
\label{28}
\<^-E|\psi^G\>=i\sqrt{\Gamma/2\pi}\frac{1}{E-(E_R-i\frac{\Gamma}{2})}\,,
\,\,-\infty_{II}<E<+\infty\,.
\end{equation}
The variable $E$ extends over the physical values (upper rim
of positive real axis first sheet = lower rim of
positive real axis second sheet) and from $-\infty_{II}$ to $0$ 
in the second sheet.
This is an idealized Breit-Wigner in contrast to the standard
Breit-Wigner for which $0\leq E<\infty$.

The results~\eqref{26} and~\eqref{27} are derived 
from the pole term definition using
the properties of the Hardy spaces~\eqref{25}~\cite{4,14}.

If there are $N$ resonances in the system,
each occurring as a pole of the $j$-th partial $S$-matrix at the positions
$z_{R_{i}}=E_{R_{i}}-i\Gamma_i/2$, then one obtains $N$
Gamow vectors $\psi_i^G$.

The Gamow vectors $\psi_i^G$ are members of a ``complex''
basis vector expansion~\cite{14}. In place of the
well known Dirac basis system expansion~\eqref{6b}
given for the Hamiltonian $H$ by
\begin{equation}
\label{a1}
\phi^+=\int_0^{+\infty}dE|E^+\>\<^+E|\phi^+\>
\end{equation}
(where a discrete sum over bound states has been ignored),
every state vector $\phi^+\in \Phi_-$ can be expanded
as
\begin{equation}
\phi^+=-\sum_{i=1}^{N}|\psi^G_i\>\<\psi_i^G|\phi^+\>
+\int_0^{-\infty_{II}}dE|E^+\>\<^+E|\phi^+\>
\label{a2}
\end{equation}
(where $-\infty_{II}$ indicates that the integration along
the negative real axis (or other contours)
is in the second Riemann sheet of the $S$-matrix).
The ``complex'' basis vector expansion~\eqref{a2}
is rigorous. This allows us to mathematically isolate
the exponentially decaying states $\psi_i^G$. It also
allows us an easy approximation by omitting the background
integral in~\eqref{a2}, and just using
\begin{equation}
\label{a3}
\phi^+=\sum_{i=1}^{N}|\psi_i^G\>c_i\,,
\quad c_i=-\<\psi_i^G|\phi^+\>\,.
\end{equation}
Then one obtains the ``effective'' theories with finite
complex Hamiltonian. For instance, for the $K_L^0 - K_S^0$ meson
system with $N=2$,
\begin{equation}
\label{a4}
\phi^+=\psi_S^G b_S+\psi_L^G b_L
\end{equation}
and one obtains the Lee-Oehme-Yang theory~\cite{lee}.
The finite dimensional approximations~\eqref{a3} have
been successfully applied to many areas of physics, in
particular to nuclear physics~\cite{ferreira,2}, which shows
that to isolate the Gamow states can be a good approximation.

Since~\eqref{27a} implies the exponential law for 
the decay rate~\cite{7} $\dot{\cal P}_{\eta}(t)=\Gamma_{\eta}e^{-\Gamma t}$
the width $\Gamma$ of the Breit-Wigner distribution~\eqref{28}
and the lifetime fulfill the exact relation
$\tau=\frac{\hbar}{\Gamma}$. This has not been obtained
before as an exact, precisely derived relation,
though it has always been assumed on the basis of
some ``approximate derivations''~\cite{15}.

Gamow vectors are ideally suited to describe resonances
(the pair~\eqref{27a} and~\eqref{27b}) in a scattering
process or quasistable particles~\eqref{27a} that decay.
Like the Dirac-Lippmann-Schwinger kets $|E^{\pm}\>$,
from which they are constructed, they do not describe interaction-free
in- or out- asymptotic states. In a theory that allows only
asymptotic particles, they are therefore not admitted.
Gamow states have all the properties that heuristically the unstable
states need to possess. In addition and unintended they give
rise to an asymmetric time evolution, which may not
have been wanted but is in agreement with 
the empirical principle of causality.
\section*{Acknowledgement}
We gratefully acknowledge valuable support of the 
Welch Foundation for the preparation of this paper
and kind hospitality of Rui Vilela Mendes in Lisbon.
 
\end{document}